\documentclass[a4paper,12pt]{article}
\usepackage{natbib}
\usepackage[english]{babel}
\renewcommand{\baselinestretch}{1.5}
\newcommand{\utwi}[1]{\mbox{\boldmath $ #1$}}

\usepackage{booktabs,dcolumn}
\usepackage{psfrag,graphicx}
\usepackage{eso-pic,graphicx}
\usepackage{amsmath}
\usepackage{graphicx}
\usepackage{pdflscape}
\usepackage{longtable}
\usepackage{epstopdf}
\usepackage{appendix}
\usepackage{dashbox}
\usepackage{algorithm,algcompatible,amsmath}
\algnewcommand\INPUT{\item[\textbf{Input:}]}%
\algnewcommand\OUTPUT{\item[\textbf{Output:}]}%

\newcolumntype{z}[1]{D{.}{.}{#1}}

\date{}
\renewcommand{\baselinestretch}{1.5}
\begin{document}

\title{
\begin{center} {\Large \bf A Bayesian realized threshold measurement GARCH framework for financial tail risk forecasting} \end{center}}

\author{Chao Wang\footnote{Corresponding author. Email: chao.wang@sydney.edu.au.}, Richard Gerlach  
\\
Discipline of Business Analytics, The University of Sydney} 
\date{} \maketitle

\begin{abstract}
\noindent
This paper proposes an innovative threshold measurement equation to be employed in a Realized-GARCH framework. The proposed framework incorporates a nonlinear threshold regression specification to consider the leverage effect and model the contemporaneous dependence between the observed realized measure and hidden volatility. A Bayesian Markov Chain Monte Carlo method is adapted and employed for model estimation, with its validity assessed via a simulation study. The validity of incorporating the proposed measurement equation in Realized-GARCH type models is evaluated via an empirical study, forecasting the 1\% and 2.5\% Value-at-Risk and Expected Shortfall on six market indices with two different out-of-sample sizes. The proposed framework is shown to be capable of producing competitive tail risk forecasting results in comparison to the GARCH and Realized-GARCH type models.
\vspace{0.5cm}

\end{abstract}

\newpage
\pagenumbering{arabic}

{\centering
\section{Introduction}
\par
}
\noindent
A major concern for financial institutions and regulators is financial risk management and forecasting. Value-at-Risk (VaR) is one of the most commonly used risk measures and employed by many financial institutions as an important risk management tool. VaR represents the market risk as one number and has become a standard risk measurement metric in recent decades. Let $\mathcal{I}_t$ be the information available at time $t$ and
\[
F_{t}(r)=Pr(r_{t}\leq r | \mathcal{I}_{t-1})
\]
be the Cumulative Distribution Function (CDF) of return $r_{t}$ conditional on $\mathcal{I}_{t-1}$. We assume that $F_{t}(.)$ is strictly increasing and continuous on the real line ${\rm I\!R}$. Under this assumption, the $\alpha$ level VaR (quantile) at time $t$ can be defined as:
\begin{equation} \label{var_def}
Q_{t}=F^{-1}_{t}(\alpha), \qquad 0 <\alpha <1. \nonumber
\end{equation}
However, critics of VaR argue that it cannot measure
expected loss in situations where there are extreme, violating returns and it is also not mathematically coherent, that is, it can favour non-diversification. Proposed by \cite{artzner1997} and \cite{artzner1999}, Expected Shortfall (ES) calculates the expected loss conditional on returns exceeding a VaR threshold and has become more widely employed in recent years for tail risk measurement. The Basel III Accord, implemented in 2019, places new emphasis on ES. Compared to VaR, ES has a number of attractive properties, including, for example, that it is a subadditive risk measure and is mathematically coherent.

Within the same framework as above for defining VaR, the $\alpha$ level ES can be shown to be equal to the tail conditional expectation of $r_t$ \citep[see][among others]{AceTas2002}:
\begin{equation}
    ES_{t}=E(r_t|r_t\leq Q_{t}, \mathcal{I}_{t-1}).
\label{e:ESdef}
\end{equation}



The Basel III Accord implemented in 2019 places new emphasis on ES. Its recommendations for market risk management are illustrated in the 2019 document \textit{Minimum Capital Requirements for Market Risk} that says: ``ES must be computed on a daily basis for the bank-wide internal models to determine market risk capital requirements. ES must also be computed on a daily basis for each trading desk that uses the internal models approach (IMA).''; ``In calculating ES, a bank must use a 97.5th percentile, one-tailed confidence level'' (Basel Committee on Banking Supervision 2019, p. 89). In our paper, we focus on the lower (left) tail forecasting and use a different sign convention for the risk measures compared to the suggested 97.5th percentile. Therefore, the $2.5\%$ probability level is studied in our paper. To widen the study, $\alpha=1\%$ is also employed.

In parametric VaR and ES estimation and forecasting, producing accurate volatility estimates and forecasts plays a crucial role. The Autoregressive Conditional
Heteroskedastic (ARCH) and Generalized ARCH (GARCH), proposed in \cite{engle1982autoregressive} and \cite{bollerslev1986generalized} respectively, have become popular in recent decades.

\cite{black1976studies} discovers the now well-known phenomenon of leverage effect (volatility asymmetry), according to which higher volatility is correlated with negative shocks in asset returns. Many studies in the literature develop asymmetric and nonlinear GARCH models to capture the leverage effect, such as EGARCH \citep{nelson1991conditional} and GJR-GARCH \citep{glosten1993relation}. \cite{poon2003forecasting} also show that asymmetric volatility models outperform symmetric models in forecasting asset return volatility.

Another popular type of volatility model developed to capture volatility asymmetry incorporates threshold autoregressive specifications, such as the one developed by \cite{tong1990non}. A double threshold GARCH framework which models the return mean and volatility asymmetry is presented in \cite{li1996double}. \cite{chen2008volatility} propose forecasting volatility using threshold heteroskedastic models by employing the intra-day high-low range.

With the availability of high frequency data, various realized measures have been proposed, such as Realized Variance (RV) (\Citealp{andersen1998answering}; \Citealp{andersen2003modeling}). These allow potentially more accurate volatility estimation compared to the daily return. \cite{hansen2012realized} propose a Realized-GARCH framework which extends the GARCH model by introducing a measurement equation that contemporaneously links the volatility and realized measure. The Realized-GARCH is shown to be capable of producing accurate volatility estimation and forecasting results, by employing RV and other realized measures. The Realized-GARCH model has gained popularity in the recent decade, with various extensions proposed.  A Realized-EGARCH is developed by \cite{hansen2016exponential} and allows multiple realized measures to be employed. Comparing to the Realized-GARCH, another development of the Realized-EGARCH model is that two leverage terms are introduced in both the GARCH and measurement equations. \cite{huang2016modeling} develop a long memory realized heterogeneous autoregressive GARCH (Realized-HAR-GARCH). \cite{chen2019bayesian} extend the Realized-GARCH model to a threshold framework by employing a threshold GARCH equation (Realized-Threshold-GARCH), so that the leverage effect can be also modelled in the GARCH equation. \cite{gerlach2022bayesian} propose a semi-parametric realized conditional autoregressive expectile framework (Realized-CARE).

For all of the above mentioned extensions of the Realized-GARCH model, a measurement equation that follows the specification in the original Realized-GARCH model is employed. In this paper, our main contribution is the proposal of an innovative threshold measurement equation that can be utilized to capture the volatility asymmetry and leverage effect. The proposed framework is motivated by the success of various threshold autoregressive volatility models. The proposed new measurement equation has the same number of parameters as the original measurement equation in the Realized-GARCH and can be employed in all the above mentioned Realized-GARCH extensions, such as Realized-EGARCH, Realized-HAR-GARCH, Realized-Threshold-GARCH and Realized-CARE. To evaluate the effectiveness of the proposed threshold measurement equation, in our paper we focus on incorporating it into the Realized-GARCH and Realized-Threshold-GARCH models. The proposed frameworks are respectively named as realized threshold measurement GARCH (Realized-T-M-GARCH) and realized double threshold GARCH (Realized-D-T-GARCH). \cite{chen2019bayesian} also consider a threshold speciﬁcation to model the conditional mean $\mu_t$ of return. As discovered in \cite{hansen2016exponential}, imposing the constraint $\mu_t=0$ (assuming $E(r_t|\mathcal{I}_{t-1})=0$) can result in better out-of-sample results in comparison to a model based on an estimated $\mu_t$. \cite{hansen2012realized} also use $\mu_t=0$ in the Realized-GARCH. Therefore, in this paper we follow the choice $\mu_t=0$ that, in practical applications, is equivalent to work with the demeaned data.


Further, we adopt an adaptive Bayesian Markov Chain Monte Carlo (MCMC) algorithm for the estimation of the proposed model. The validity of the employed MCMC is evaluated via a simulation study. To evaluate the performance of the proposed Realized-T-M-GARCH and Realized-D-T-GARCH models, the accuracy of the associated 1\% and 2.5\% VaR and ES forecasts is assessed via comprehensive empirical studies which find that the proposed model produces competitive tail risk forecasting results compared to the GARCH and Realized-GARCH type models.



The paper is structured as follows. Section \ref{model_proposed_section} reviews the Realized-GARCH model and proposes the Realized-T-M-GARCH and Realized-D-T-GARCH models. The associated
likelihood and the MCMC algorithm for model estimation are presented in Section \ref{parameter_estimation_section}.
The simulation and empirical results are discussed in Section \ref{simulation_section} and Section \ref{data_empirical_section} respectively.
Section \ref{conclusion_section} concludes the paper.

{\centering
\section{Model Proposed}\label{model_proposed_section}
\par
}
\noindent

\subsection{Realized-GARCH}
The Realized-GARCH model with log specification of \cite{hansen2012realized} can be written as:
\begin{eqnarray}\label{rgarch}
r_t&=& \sqrt{h_t} z_t, \\ \nonumber
\text{log}(h_t)&=& \omega +\beta \text{log}(h_{t-1})+ \gamma \text{log}(x_{t-1}), \\ \nonumber
\text{log}(x_t)&=& \xi +\varphi \text{log}(h_{t})+ \tau_1 z_t + \tau_2 (z_t^2-1)+  \sigma_{\varepsilon} \varepsilon_t,
\end{eqnarray}
where $r_t= 100 \times [\text{log}(C_t)-\text{log}(C_{t-1})]$ is the percentage log-return for day $t$,
$z_t \stackrel{\rm i.i.d.} {\sim} D_1(0,1)$, $ \varepsilon_t \stackrel{\rm i.i.d.} {\sim} D_2(0,1)$, $h_t$ is the conditional variance (volatility square) and $x_t$ is a realized measure, for example, RV. $D_1(0,1)$ and $D_2(0,1)$ indicate distributions that have mean 0 and variance 1.
The three equations in order in model (\ref{rgarch}) are: the return equation,
the GARCH equation and the measurement equation, respectively. The measurement equation is an observation equation that captures the dependence between the latent volatility and realized measure. The term $\tau_1 z_t + \tau_2 (z_t^2-1)$ captures the leverage effect.



The leverage term in the EGARCH model employs the form $\tau_1 z_t + \tau_2 \left (|z_t|- E[|z_t|] \right )$. Therefore, the leverage term $\tau_1 z_t + \tau_2 (z_t^2-1)$ in Realized-GARCH is a quadratic variable (constructed from Hermite polynomials) of the version in EGARCH (see \Citealp{hansen2016exponential}, p. 270, for a detailed discussion and comparison of the two specifications).

\cite{hansen2012realized} choose Gaussian errors, for example, $D_1(0,1) = D_2(0,1) \equiv N(0,1)$. \cite{watanabe2012quantile} allows $D_1(0,1)$ to be a standardized
Student's t and skew t of \cite{fernandez1998bayesian}. Student's t is also the choice of $D_1(0,1)$ in \cite{gerlach2016forecasting}. \cite{contino2017bayesian} further test $D_2(0,1)$ as a Student's t distribution, while their findings show that changing the distribution of $D_2(0,1)$ will not significantly affect the performance of the model, so $D_2(0,1)\equiv N(0,1)$.

\subsection{Realized threshold measurement GARCH}
In a regression model, it is commonly assumed that the coefficients are fixed, such as the $\text{log}(x_t)= \xi +\varphi \text{log}(h_{t})$ part in the measurement equation of the Realized-GARCH. However, in some situations it is more appropriate to allow the regression coefficients to vary as a function of time or as a function of some relevant variables. Such regression frameworks are called switching regression or regime regression models (for example, \Citealp{goldfeld1972nonlinear}; \Citealp{granger1993modelling}). The GJR-GARCH utilizes these kinds of frameworks in modelling the volatility asymmetry and leverage effect, and takes the following form for the volatility component:
\begin{equation}\label{GJR_GARCH}
h_t=\omega +\beta h_{t-1}+ (\gamma +\alpha I(r_{t-1} \le 0 )) r_{t-1}^2, \\
\end{equation}
where $I(A)$ is the indicator function taking value 1 if event $A$ occurs and 0 otherwise. Therefore, the framework is capable of modelling the volatility asymmetrically according to whether the lagged return is positive or negative.

Another popular way of capturing the volatility asymmetry is by incorporating a threshold GARCH specification \citep[see, for example,][among others]{li1996double}:
\begin{eqnarray}\label{T_GARCH}
 h_t=
\begin{cases}
   \omega_1 +\beta_1 h_{t-1}+ \gamma_1 r_{t-1}^2, &  \zeta_{t-1} \leq c, \\
   \omega_2 +\beta_2 h_{t-1}+ \gamma_2 r_{t-1}^2, &  \zeta_{t-1} > c,
\end{cases}
\end{eqnarray}
where $\zeta_{t}$ is the threshold variable and $c$ is the threshold value.

Motivated by specifications (\ref{GJR_GARCH}) and (\ref{T_GARCH}), we propose an innovative threshold measurement equation to be employed in the Realized-GARCH framework. The proposed framework is named as realized threshold measurement GARCH (Realized-T-M-GARCH):

\noindent
\textbf{Realized-T-M-GARCH}
\begin{eqnarray}\label{r_t_m_garch}
r_t&=& \sqrt{h_t} z_t, \\ \nonumber
\text{log}(h_t)&=& \omega +\beta \text{log}(h_{t-1})+ \gamma \text{log}(x_{t-1}), \\ \nonumber
m_t&=&
    \begin{cases}
         \xi_1 +\varphi_1 \text{log}(h_{t}) , &  \zeta_{t} \leq c, \\
         \xi_2 +\varphi_2 \text{log}(h_{t}) , &  \zeta_{t} > c,
    \end{cases} \\ \nonumber
\text{log}(x_{t})&=& m_t + \sigma_{\varepsilon} \varepsilon_t .\nonumber
\end{eqnarray}


Compared to the Realized-GARCH as in model (\ref{rgarch}), the proposed Realized-T-M-GARCH models the leverage effect in a different (threshold) manner. In addition, the proposed threshold measurement equation has the same number of parameters as the one in the Realized-GARCH. Further, as discussed the proposed threshold measurement equation can be conveniently employed in other extensions of the Realized-GARCH framework, e.g., the Realized-Threshold-GARCH (Realized-T-GARCH) developed by \cite{chen2019bayesian}. The specification of Realized-T-GARCH is shown below. As can be seen, it employs a threshold specification for the GARCH equation, while a measurement equation in the original form of the Realized-GARCH model is used.

\noindent
\textbf{Realized-T-GARCH}
\begin{eqnarray}\label{r_d_t_garch}
r_t&=& \sqrt{h_t} z_t, \\ \nonumber
\text{log}(h_t)&=&
    \begin{cases}
        \omega_1 +\beta_1 \text{log}(h_{t-1})+ \gamma_1 \text{log}(x_{t-1}), &  \zeta_{t-1} \leq c,  \\ \nonumber
        \omega_2 +\beta_2 \text{log}(h_{t-1})+ \gamma_2 \text{log}(x_{t-1}) &  \zeta_{t-1} > c,
    \end{cases} \\ \nonumber
\text{log}(x_t)&=& \xi +\varphi \text{log}(h_{t})+ \tau_1 z_t + \tau_2 (z_t^2-1)+  \sigma_{\varepsilon} \varepsilon_t.
\end{eqnarray}

Therefore, incorporating the proposed threshold measurement equation into the Realized-Threshold-GARCH of \cite{chen2019bayesian}, a realized double threshold GARCH (Realized-D-T-GARCH) framework can be proposed as:

\noindent
\textbf{Realized-D-T-GARCH}
\begin{eqnarray}\label{r_d_t_garch}
r_t&=& \sqrt{h_t} z_t, \\ \nonumber
\text{log}(h_t)&=&
    \begin{cases}
        \omega_1 +\beta_1 \text{log}(h_{t-1})+ \gamma_1 \text{log}(x_{t-1}), &  \zeta_{t-1} \leq c,  \\ \nonumber
        \omega_2 +\beta_2 \text{log}(h_{t-1})+ \gamma_2 \text{log}(x_{t-1}) &  \zeta_{t-1} > c,
    \end{cases} \\ \nonumber
m_t&=&
    \begin{cases}
         \xi_1 +\varphi_1 \text{log}(h_{t}) , &  \zeta_{t} \leq c, \\
         \xi_2 +\varphi_2 \text{log}(h_{t}) , &  \zeta_{t} > c,
    \end{cases} \\ \nonumber
\text{log}(x_{t})&=& m_t + \sigma_{\varepsilon} \varepsilon_t. \nonumber
\end{eqnarray}


In our paper, we choose the threshold variable $\zeta_{t}$ to be self-exciting, that is, $\zeta_{t}= r_{t}$, and the threshold value $c=0$, as typical choices in the literature. Therefore, the proposed threshold measure equations can be used to capture the leverage effect. Its properties are discussed and compared with the original measurement equation in Realized-GARCH in Section \ref{parameter_estimates_section}. Although not investigated in our paper, the proposed threshold measurement equation has more flexibility than the one in the Realized-GARCH and can be easily further extended. For example, the threshold variable $\zeta_{t}$ can be chosen as the realized measure $x_t$ and threshold value $c$ can be estimated instead of fixed, thus the size asymmetry can be also considered. The threshold variable $\zeta_{t}$ can also be selected as other exogenous economic variables to allow for a potentially more flexible and informative dynamic process.

Stationarity is an important issue in volatility modelling. As derived in \cite{hansen2012realized}, by substituting the threshold measurement equation into the volatility equation the
required stationary condition for the Realized-T-M-GARCH model is:
\begin{equation}\label{stationarity_condition}
\beta + \gamma \varphi_1<1; \beta + \gamma \varphi_2<1.
\end{equation}

Similarly, the stationarity condition for the Realized-D-T-GARCH model can be shown as:
\begin{equation}\label{stationarity_condition_dt}
\beta_1 + \gamma_1 \varphi_1<1; \beta_2 + \gamma_2 \varphi_2<1.
\end{equation}

Since the log specification is used, the non-negativity concern associated with volatility is not an issue for the Realized-T-M-GARCH and Realized-D-T-GARCH models. In the empirical section, the performance of the proposed Realized-T-M-GARCH and Realized-D-T-GARCH models will be compared with the Realized-GARCH and Realized-T-GARCH. Following \cite{watanabe2012quantile} and \cite{gerlach2016forecasting}, we focus on testing Student's t distribution as the distribution $D_1$ in the return equation, although other distributions, such as skew t distribution of \cite{hansen1994autoregressive}, could be also employed and tested.


{\centering
\section{Likelihood and Bayesian Estimation}\label{parameter_estimation_section}
\par
}
\noindent
\subsection{Likelihood}
As in \cite{hansen2012realized}, when $D_1 = D_2 \equiv N(0,1)$ the log-likelihood function for model (\ref{rgarch}) is:
\begin{equation}\label{RGGG_lik}
\ell (\utwi{r},\utwi{x};\utwi{\theta})=\underbrace{ -\frac {1}{2} \sum_{t=1}^{n} \left[ \text{log}(2 \pi)+\text{log}(h_t)+ r_t^2 / h_t \right]}_{\ell (\utwi{r};\utwi{\theta})}
 \underbrace{-\frac {1}{2} \sum_{t=1}^{n} \left[ \text{log}(2 \pi)+\text{log}(\sigma_{\varepsilon}^2)+ \epsilon_t^2/\sigma_{\varepsilon}^2
 \right]}_{\ell (\utwi{x}|\utwi{r};\utwi{\theta})},
\end{equation}
where $\utwi{r}= \{r_1, r_2,\ldots, r_n \}$, $\utwi{x}= \{x_1, x_2,\ldots, x_n \}$ and $n$ is the in-sample size. $\utwi{\theta}$ represents the parameter vector, and $\epsilon_t= \text{log}(x_t)-\xi -\varphi \text{log}(h_{t})-\tau_1 z_t - \tau_2 (z_t^2-1)$. The log-likelihood $\ell (\utwi{r},\utwi{x};\utwi{\theta})$ function equals the sum of two parts $\ell (\utwi{r};\utwi{\theta})$ and $\ell (\utwi{x}|\utwi{r};\utwi{\theta})$, which are derived from the GARCH and measurement equation respectively.

In our paper, we test the proposed Realized-T-M-GARCH framework via employing return equation error as $D_1 \equiv t_{\nu}(0,1)$ and measurement equation error as $D_2 \equiv N(0,1)$. $t_{\nu}(0,1)$ represents a Student's t distribution with $\nu$ degrees of freedom and variance scaled to 1 (by using $\sqrt{\frac{\nu-2}{\nu}}$ factor). The framework is called Realized-T-M-GARCH-tG.

$\ell (\utwi{x}|\utwi{r};\utwi{\theta})$ remains the same as the one in Equation (\ref{RGGG_lik}) under the threshold measurement equation specification, as long as the $D_2$ distribution remain unchanged. Therefore, by updating $\ell (\utwi{r};\utwi{\theta})$ according to the Student's t distribution, the log-likelihood function of the proposed Realized-T-M-GARCH-tG model is:
\begin{eqnarray}\label{r_t_m_garch_lik}
\ell (\utwi{r},\utwi{x};\utwi{\theta}) &=& \underbrace{ - \sum_{t=1}^{n} \left[ A(\nu)+ \frac{1}{2}\log(h_t)+ \frac{\nu+1}{2} \log \left(1+ \frac{r_t^2}{h_t (\nu-2)} \right) \right]}_{\ell (\utwi{r};\utwi{\theta})}  \\ \nonumber
  && \underbrace{-\frac {1}{2} \sum_{t=1}^{n} \left[ \log(2 \pi)+\log(\sigma_{\varepsilon}^2)+ \epsilon_t^2/\sigma_{\varepsilon}^2
 \right]}_{\ell (\utwi{x}|\utwi{r};\utwi{\theta})}
\end{eqnarray}
where $\epsilon_t=\text{log}(x_t)-\xi_1 -\varphi_1 \text{log}(h_{t})$ when $r_{t} \le 0$, $\epsilon_t=\text{log}(x_t)-\xi_2 -\varphi_2 \text{log}(h_{t})$ when $r_{t} > 0$, and
$A(\nu) = - \log\left(\Gamma\left(\frac{\nu+1}{2}\right) \right) + \frac{1}{2}\log(\pi(\nu-2))  + \log\left(\Gamma \left(\frac{\nu}{2}\right) \right) $. The parameter vector to be
estimated is $\utwi{\theta}=(\omega,\beta,\gamma,\xi_1,\varphi_1,\xi_2,\varphi_2,\sigma_{\varepsilon},\nu)^{'}$, under constraints in Equation (\ref{stationarity_condition}). $\nu > 4$ is further restricted to ensure the
first four moments of the return error distribution are finite.

For comparison purposes, we also incorporate Student's t distribution as $D_1$ and Gaussian distribution as $D_2$ in the Realized-GARCH, Realized-T-GARCH and Realized-D-T-GARCH models. For these models, the log-likelihood is identical to the Realized-T-M-GARCH framework, as the $D_1$ and $D_2$ distribution remain unchanged.


\subsection{Bayesian Estimation}\label{bayesian_estimation_section}
\noindent
Motivated by the MCMC results in \cite{gerlach2016forecasting} and \cite{gerlach2022bayesian}, an adaptive MCMC procedure is employed for the estimation of Realized-GARCH, Realized-T-GARCH and the proposed Realized-T-M-GARCH and Realized-D-T-GARCH models. This also aims to make the later model performance comparison among these four models a fair one.



The motivation of employing the Bayesian MCMC approach (instead of maximum likelihood estimation (MLE)) is discussed in \cite{gerlach2016forecasting}. For example, the estimation of Realized-T-M-GARCH requires constrained MLE to ensure stationarity, and this may cause issues in the optimization and in the standard error calculation \citep{silvapulle2005constrained}. Therefore, in this paper we have adopted and extended the adaptive MCMC method in \cite{gerlach2022bayesian}, using the following two steps: a burn-in step and an ``independent'' Metropolis-Hastings (IMH) step.

First, parameter blocking is employed for the MCMC estimation of Realized-T-M-GARCH-tG. Three blocks are chosen as: $\utwi{\theta_1}=(\omega,\beta,\gamma, \varphi_1, \varphi_2)^{'}$; $\utwi{\theta_2}=(\xi_1, \xi_2,\sigma_{\varepsilon})^{'}$
and $\utwi{\theta_3}=(\nu)$. For the proposed double threshold framework, the first parameter block is $\utwi{\theta_1}=(\omega_1,\beta_1,\gamma_1,\omega_2,\beta_2,\gamma_2, \varphi_1, \varphi_2)^{'}$, $\utwi{\theta_2}=(\xi_1, \xi_2,\sigma_{\varepsilon})^{'}$  and $\utwi{\theta_3}=(\nu)$. The choice is motivated by the fact that parameters within the same block are more strongly correlated, in the posterior (or likelihood), than those between blocks. For example, the stationarity condition of Realized-T-M-GARCH causes correlation between iterates of $\beta,\gamma,\varphi_1, \varphi_2$, thus they are kept together in one block.

Similarly, three blocks for the Realized-GARCH-tG model are also used:  $\utwi{\theta_1}=(\omega,\beta,\gamma, \varphi)^{'}$; $\utwi{\theta_2}=(\xi,\tau_1, \tau_2,\sigma_{\varepsilon})^{'}$ and $\utwi{\theta_3}=(\nu)$. For the Realized-T-GARCH-tG framework, the first parameter block is $\utwi{\theta_1}=(\omega_1,\beta_1,\gamma_1,\omega_2,\beta_2,\gamma_2, \varphi)^{'}$, and $\utwi{\theta_2}$ and $\utwi{\theta_3}$ remain the same. Uninformative priors are chosen for both models over the possible stationarity region, that is, $\pi(\utwi{\theta})\propto I(A)$, which is a flat prior for $\utwi{\theta}$ over the region $A$ satisfying $\nu > 4$ and the stationarity conditions, e.g., Equations (\ref{stationarity_condition}) and (\ref{stationarity_condition_dt}).

For the burn-in period,  a Metropolis algorithm (Metropolis \emph{et al.}, 1953) employing a mixture of 3 Gaussian proposal distributions, with a random walk mean vector, is utilized for each block of parameters. In addition, an iterative ``epoch'' method, as in \cite{chen2017dynamic}, is employed, with the aim of enhancing the convergence of MCMC chains. The proposal variance--covariance (var--cov) matrix of each block in each mixture element is set as $C_i \Sigma_1$, where
$C_1 =1;C_2 =100;C_3 =0.01$, with $\Sigma_1$ initially set to $\frac{2.38}{\sqrt{(d_i)}}I_{d_i}$, where $d_i$ is the dimension of the
block ($i$) of parameters being generated and $I_{d_i}$ is the identity matrix of dimension $d_i$. This proposal var--cov matrix
is subsequently tuned with the aim of meeting a target acceptance rate of $23.4\%$ (if $d_i>4$, or $35\%$ if $2 \le d_i \le 4$, or $44\%$ if
$d_i=1$), as standard, via the algorithm used in \cite{gelman1997weak}. After running the first epoch with 20,000 iterations, the var--cov matrix of each parameter block is calculated after discarding the first 2000 iterates. This updated var--cov matrix is then employed in the proposal distribution of the next epoch. After running each epoch, the standard deviations of each parameter chain are also calculated and compared to that of the previous epoch. The epoch process is repeated until the mean absolute percentage change of these standard deviations is less than 10\%, which takes approximately three to four epochs in both simulation and empirical studies.

In the IMH step (10,000 iterations), again a mixture of three Gaussian proposal distributions is utilized for each parameter block. The sample mean vector of the last epoch iterates (after discarding the first 2000 iterates) in the burn-in period is used as the mean vector of the IMH step. For the var--cov matrix, after discarding the first 2000 iterates, the sample covariance matrix of the last epoch iterates for each block is calculated as $\Sigma_2$. Then the proposal  var--cov matrix in each element is calculated as $C_i \Sigma_2$, where $C_1 =1;C_2 =100;C_3 =0.01$.

Lastly, all the IMH iterates (still discarding the first 2000 iterates) are employed to calculate the VaR and ES forecasts, the posterior means of which are used as the final tail risk forecasts.

{\centering
\section{Simulation Study}\label{simulation_section}
\par
}
\noindent
A simulation study is designed to illustrate the validity of the adapted MCMC
in terms of parameter estimation and VaR and ES forecasting accuracy of the proposed models. To limit the focus of the study, the simulation section considers the Realized-T-M-GARCH-tG model.

1000 replicated datasets of size $n=1900$ (chosen based on empirical study in-sample size) are simulated from the following simulation model which follows the Realized-T-M-GARCH-tG specification.

\textbf{Simulation Model}
\begin{eqnarray}\label{rgarch_sim}
r_t&=& \sqrt{h_t} z_t, \\ \nonumber
\text{log}(h_t)&=& 0.1 + 0.65 \text{log}(h_{t-1})+ 0.3 \text{log}(x_{t-1}), \\ \nonumber
m_t&=&
    \begin{cases}
         -0.2 + 0.92 \text{log}(h_{t}) , &  r_{t} \leq 0, \\
         -0.5 + 0.95 \text{log}(h_{t}) , &  r_{t} > 0,
    \end{cases} \\ \nonumber
\text{log}(x_{t})&=& m_t + 0.6 \varepsilon_t \, , \nonumber
\end{eqnarray}

where $z_t \stackrel{\rm i.i.d.} {\sim} t_{\nu=10}(0,1)$ and $\varepsilon_t \stackrel{\rm i.i.d.} {\sim} N(0,1)$. The true values of parameters in the simulation model are selected based on the typical parameter estimates in the empirical study (to be shown in detail in Section \ref{parameter_estimates_section}). In the simulation model, $r_{t}$ is analogous to the daily return and $x_t$ is analogous to the daily realized measure.

The ``True'' one-step-ahead $\alpha$ level VaR forecast from the above simulation model is calculated as:
\begin{equation}\label{var_t_dist_1}
Q_{t+1}=  \sqrt{h_{t+1}} t_{\nu}^{-1}(\alpha)\sqrt{\frac{\nu-2}{\nu}}, \nonumber
\end{equation}
where $t_{\nu}^{-1}(\alpha)$ is the inverse of Student's t CDF with the $\nu$ degrees of freedom on probability level $\alpha$. The ES forecast from the same model is calculated as:
\begin{equation}\label{es_t_dist_1}
\text{ES}_{t+1}=  - \sqrt{h_{t+1}}  \left( \frac{g_{\nu}(t_{\nu}^{-1}(\alpha))}{\alpha} \right) \left( \frac{\nu+ (t_{\nu}^{-1}(\alpha))^{2}}{\nu-1} \right) \sqrt{\frac{\nu-2}{\nu}}, \nonumber
\end{equation}
where $g_{\nu}(.)$ is the Student's t Probability Density Function (PDF).

These ``True'' VaR and ES forecasts are calculated for each dataset. The averages of the these ``True'' forecasts, over the 1000 datasets, are given in the ``True'' column of Table \ref{simu_stat_1}.

Simulation results of both 1\% and 2.5\% probability levels are presented in Table \ref{simu_stat_1}. All initial parameter values are arbitrarily set equal to $0.25$ to start the MCMC chains. The Mean column shows the average parameter estimates and average VaR and ES forecasts across 1000 simulated datasets. The Root Mean Squared Error (RMSE) between the parameter estimates and parameter true values (and VaR and ES forecasts and their ``True'' values) are also shown. In general, the MCMC algorithm produces accurate parameter estimates and tail-risk forecasts in terms of bias (difference between True and Mean columns) and precision (RMSE). In particular, both VaR and ES forecasts on 1\% and 2.5\% levels are close to their true values with bias less than 0.03 and RMSE less than 0.14. This means the employed MCMC procedure could produce accurate parameter estimates and tail risk forecasts.



\renewcommand{\baselinestretch}{1.1}
\begin{table}[!ht]
\begin{center}
\caption{\label{simu_stat_1} \small Summary statistics for the MCMC estimator of the Realized-T-M-GARCH-tG model with the simulated datasets.}\tabcolsep=10pt
\begin{tabular}{lcccccccccc} \hline
$n=1900$    &        &\multicolumn{2}{c}{MCMC}      \\
Parameter   &  True  &      Mean    &       RMSE    \\ \hline
$\omega$&	 0.1000 &  0.0993 &	 0.0181 \\
$\beta$&	 0.6500 &	 0.6443 &	 0.0231   \\
$\gamma$&	 0.3000 &	  0.3001 &	 0.0273  \\
$\xi_1$&	-0.2000 &	-0.2018 &	 0.0443 \\
$\varphi_1$&	 0.9200 & 0.9369 &	 0.0842   \\
$\xi_2$&	-0.5000 &	 -0.4971 &	 0.0448 \\
$\varphi_2$&	 0.9500 & 0.9630 &	 0.0868 	  \\
$ \sigma_{\varepsilon} $&	 0.6000 &0.6011 &	 0.0100 	\\
$\nu$&	 10.0000 & 12.0750 	& 3.7380  	 \\
1\% $Q_{t+1}$	&-2.4576 &-2.4423 &	 0.0889 	\\
2.5\% $Q_{t+1}$&	-1.9813& -1.9745 &	 0.0632  \\
$\text{1\% ES}_{t+1}$&	-2.9907 &-2.9622 	& 0.1383 	 \\
$\text{2.5\% ES}_{t+1}$&	-2.5068 &	-2.4892 &	 0.0967  \\

\hline
\end{tabular}
\end{center}
\end{table}

\begin{figure}[htp]
     \centering
\includegraphics[width=.9\textwidth]{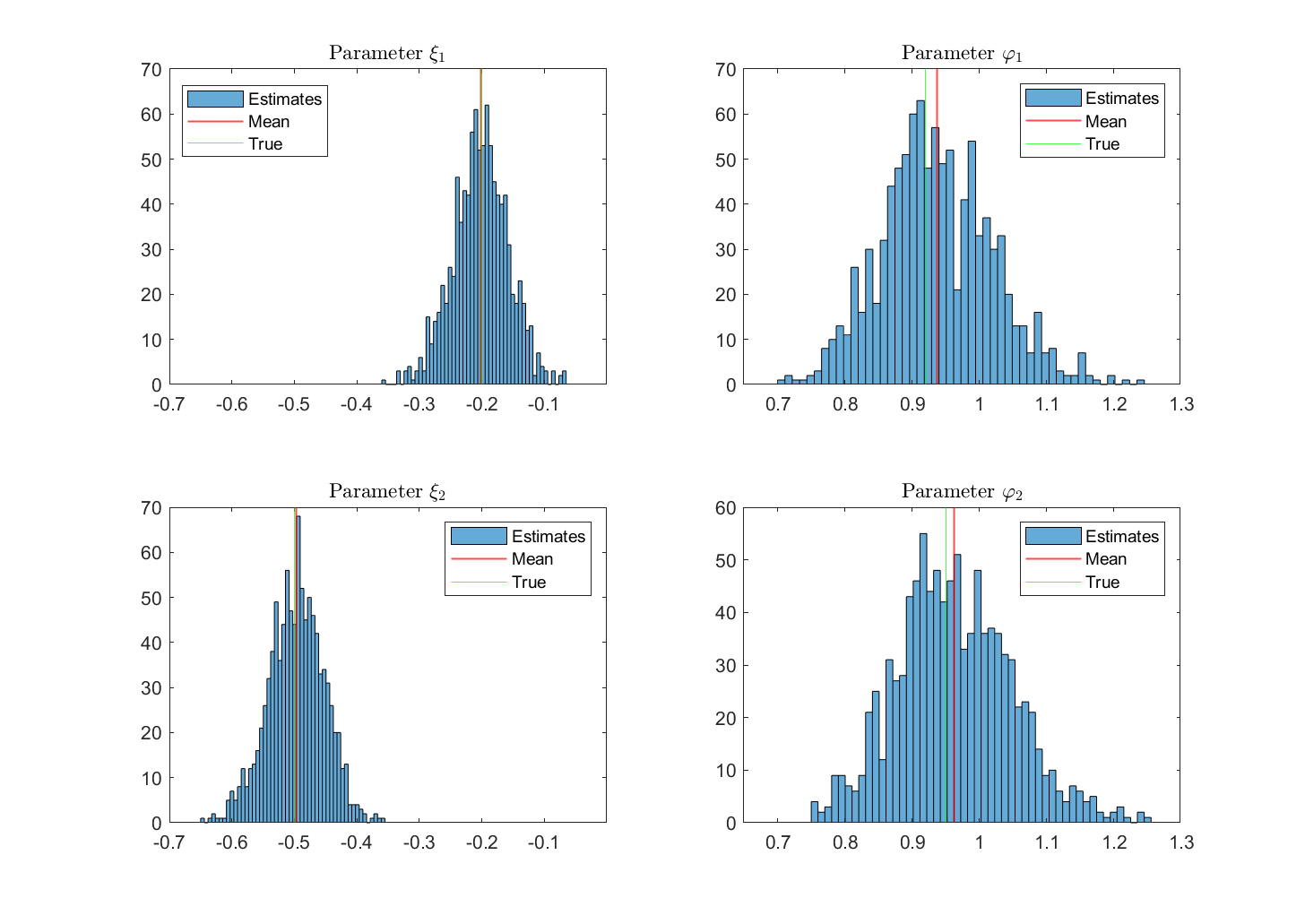}
\caption{\label{sim_para_estimates} Histograms of 1000 parameter estimates for $\xi_1$\&$\xi_2$ and $\varphi_1$\&$\varphi_2$ in the threshold measurement equation. True vertical line represents parameter true value from the simulation model. Mean vertical line represents the average of 1000 parameter estimates from MCMC.}
\end{figure}

Further, since the main development of this paper is the threshold measurement equation, Figure \ref{sim_para_estimates} presents the histograms of the parameter estimates of all 1000 simulated datasets for $\xi_1$\&$\xi_2$ and $\varphi_1$\&$\varphi_2$. As can be seen, although parameters $\xi_1$\&$\xi_2$ and $\varphi_1$\&$\varphi_2$ have different true values, the adapted MCMC algorithm is capable of estimating all four parameters accurately, with the RMSE values around 0.04 for $\xi_1$\&$\xi_2$ and around 0.08 for $\varphi_1$\&$\varphi_2$ respectively.

{\centering
\section{Empirical Study}\label{data_empirical_section}
\par
}
\noindent
\subsection{Data and forecasting setup}\label{data_description_section}
\noindent
Six market indices are assessed in the empirical section, including S\&P500, NASDAQ (both US), FTSE100 (UK), DAX (Germany), SMI (Switzerland), and ASX200 (Australia), in the time period from April 2000 to December 2015. The high frequency closing prices, observed at 5-minute intervals within trading hours, are downloaded from Thomson Reuters Tick History. The 5-minute data are employed to calculate the daily RV. The daily closing prices are also collected and used to calculate the daily return.


As discussed in Section \ref{simulation_section}, the daily returns employed are close-to-close, including overnight price movement, while the RV is only calculated while the market is open. Therefore, the employed realized measures may have some downside bias for the true volatility of close-to-close returns, which can be captured and corrected by the $\xi_1$ and $\xi_2$ parameters in the proposed threshold measurement framework.

A fixed size in-sample dataset is employed for estimation, combined with a rolling window approach, to produce each one-step-ahead VaR and ES forecasts on 1\% and 2.5\% probability levels. Table \ref{t:quantile_loss_gfc} reports the in-sample size for each series, which differs due to different non-trading days occurred in each market.

Two forecasting studies with different out-of-sample sizes are conducted. The first study aims to assess the performance of models for the 2008 Global Financial Crisis (GFC) period, thus the initial date of the out-of-sample forecasting period is chosen as January 2008. Then for each index the out-of-sample size $m$ is chosen as 400, meaning that the end of the forecasting period is around August 2009.

An eight year out-of-sample period is employed in the second forecasting study with the out-of-sample start date still chosen as January 2008 and out-of-sample size $m$ as 2000. Therefore, the end of the forecasting period is around December 2015.

The Realized-GARCH, Realized-T-GARCH and proposed Realized-T-M-GARCH and Realized-D-T-GARCH models, with Student's t return error, all estimated with MCMC, are included in the forecasting study to compare their performance.


In addition, for comparison purposes, EGARCH and GJR-GARCH, both with Student's t distribution, are also included. We also include a semi-parametric filtered historical simulation approach. More specifically, the series of in-sample conditional volatility $\sqrt{\hat{h}_t}$ is estimated based on the fitted EGARCH-t model. The error quantiles and tail expectations are then estimated by computing the empirical $\alpha$ level quantile ($\hat{q}{(\alpha)}$) and empirical $\alpha$ level tail average ($\hat{c}{(\alpha)}$) of the standardized returns $r_t / \sqrt{\hat{h}_t}$. Finally, the $\alpha$ level VaR and ES forecasts are obtained by multiplying $\hat{q}{(\alpha)}$ and $\hat{c}{(\alpha)}$, respectively, by the volatility forecast $\sqrt{\hat{h}_{t+1}}$ from the fitted EGARCH-t model. The model is named as EGARCH-t-HS. In addition to the EGARCH-t model, we also test employing GJR-GARCH-t in the historical simulation approach (GJR-GARCH-t-HS). All these GARCH type models are estimated by MLE.


%

\subsection{Parameter estimates}\label{parameter_estimates_section}
\par

In this section, we study the parameter estimates from the proposed models and their comparison to the Realized-GARCH and Realized-T-GARCH models. In particular, we investigate how the parameters in the proposed threshold measurement equation behave and how the leverage effect can be successfully captured.

\subsubsection{One forecasting step results}

For the four competing Realized-GARCH type models with the Student's t return error, Table \ref{parameter_estimates} shows the parameter posterior means and the lower and upper quantiles (LQ and UQ) of the 95\% credible intervals (CI), using the first moving window of S\&P 500 returns. We have the following observations.

\begin{table}[!ht]
\begin{center}\footnotesize
\caption{\label{parameter_estimates} \small Parameter posterior means and lower and upper quantiles of the 95\% credible intervals of four Realized-GARCH type models with the Student's t return error, using the first moving window of S\&P 500 returns.}\tabcolsep=10pt
\begin{tabular}{lccc|lcccccc} \hline
    &        \multicolumn{3}{c}{Realized-GARCH-tG} &  &\multicolumn{3}{c}{Realized-T-M-GARCH-tG}      \\
 Parameter                          & Mean    & LQ     & UQ      &   Parameter                         & Mean    & LQ      & UQ      \\ \hline
$\omega$                   & 0.1007  & 0.0749  & 0.1268  & $\omega$                   & 0.1018	&0.0755	&0.1319\\
$\beta$                    & 0.6744  & 0.6344  & 0.7151  & $\beta$                    & 0.6898&	0.6467&	0.7330  \\
$\gamma$                   & 0.2997  & 0.2573  & 0.3420   & $\gamma$                   & 0.3013&	0.2567&	0.3433   \\
$\xi$                       & -0.3493 & -0.4087 & -0.2899 & $\xi_1$                     &-0.2562&	-0.3344	&-0.1831 \\
$\varphi$                      & 1.0069  & 0.9353  & 1.0893  & $\varphi_1$                    & 0.9325&	0.8521&	1.0126\\
$\tau_1$                    & -0.0632 & -0.0869 & -0.0389 & $\xi_2$                     &  -0.4349	&-0.5069&	-0.3606\\
$\tau_2$                    & 0.1076  & 0.0935  & 0.1235  & $\varphi_2$                    &  0.9743&	0.8862&	1.0620 \\
$\sigma_{\varepsilon}   $  & 0.5101  & 0.4941  & 0.5269  & $\sigma_{\varepsilon}   $  &   0.5419&	0.5251	&0.5599\\
$\nu$ &	17.5772	 &10.0084 &	28.6255	 &$\nu$ &17.5017&	10.1196&	28.2448	\\ \hline

    &        \multicolumn{3}{c}{Realized-T-GARCH-tG} &  &\multicolumn{3}{c}{Realized-D-T-GARCH-tG}      \\
 Parameter                          & Mean    & LQ      & UQ      &   Parameter                         & Mean    & LQ      & UQ      \\ \hline

$\omega_1$                 & 0.2139  & 0.1782  & 0.2525  & $\omega_1$                 &  0.2056&	0.1750 &	0.2442\\
$\beta_1$                  & 0.7019  & 0.6534  & 0.7533  & $\beta_1$                  &  0.7090&	0.6597&	0.7570 \\
$\gamma_1$ & 0.2951  & 0.2471  & 0.3418  & $\gamma_1$ & 0.2997&	0.2527&	0.3495 \\
$\omega_2$                 & -0.0549 & -0.0822 & -0.0226 & $\omega_2$                 &  -0.0609	&-0.0863&	-0.0325\\
$\beta_2$                  & 0.7654  & 0.7173  & 0.8114  & $\beta_2$                  &   0.7911&	0.7400&	0.8350\\
$\gamma_2$                 & 0.1953  & 0.1536  & 0.2378  & $\gamma_2$                 &0.1724&	0.1278	&0.2172  \\
$\xi$                       & -0.3450  & -0.4123 & -0.2863 & $\xi_1$                     & -0.2317&	-0.2989	&-0.1594 \\
$\varphi$                      & 0.9678  & 0.9056  & 1.0293  & $\varphi_1$                    &  0.9377	&0.8591&	1.0377\\
$\tau_1$                    & -0.0708 & -0.0932 & -0.0493 & $\xi_2$                     & -0.4287&	-0.5000&	-0.3490 \\
$\tau_2$                    & 0.1105  & 0.0963  & 0.1268  & $\varphi_2$                    & 0.9559&	0.8757	&1.0532  \\
$\sigma_{\varepsilon}   $  & 0.4929  & 0.4785  & 0.5090   & $\sigma_{\varepsilon}   $  &0.5234	&0.5064	&0.5409\\
$\nu$&	19.5104&	10.8958	&29.1515&	$\nu$&19.3094&	11.2353&	29.1665	\\

\hline
\end{tabular}
\end{center}
\end{table}

Overall, the estimated GARCH equation parameter values of the proposed Realized-T-M-GARCH and Realized-D-T-GARCH models, in general, are consistent with those in the Realized-GARCH and Realized-T-GARCH models, while distinctive behaviours are observed for parameters in the measurement equation.

First, regarding the GARCH equation, in a Realized-GARCH the $\gamma$ parameter (coefficient of lagged realized measure) is typically estimated to be between 0.3 and 0.55 in the empirical study. This parameter may be compared with $\alpha$ in a conventional GARCH model, which measures the coefficient associated with the conditional variance estimator (squared return in GARCH). With a more efficient and informative realized measure employed, the estimated $\gamma$ parameter is, in general, greater than the estimated $\alpha$ in GARCH. Under the Realized-T-M-GARCH framework, we have similar observations.

Second, regarding the measurement equation, in a Realized-GARCH, estimates of $\varphi$ in the measurement equation are close to unity, which suggests that the realized measure $x_t$ is roughly proportional to the conditional variance of daily returns. The $\xi$ parameter estimates in a Realized-GARCH are always negative. This suggests a negative correction (given by estimates of $\xi$) is required. One potential reason for this is that the returns employed in this paper are close-to-close and include overnight price movements, but the realized variance is only measured when the market is open and may underestimate the true volatility on average (downside bias).

On p. 749 of Chen and Watanabe (2018), they have discussed how the leverage effect can be captured by the $ \tau_1 z_t + \tau_2 (z_t^2-1)$ term in the measurement equation of Realized-GARCH. Given a $\hat{h}_t$, if $z_t\le0$ (meaning $r_t\le0$), then $\tau_1 z_t$ is a positive figure, as the $\tau_1$ estimate is always negative. Adding such positive figure into the $\xi$ of the measurement equation will produce a larger $\xi+\tau_1 z_t$ term, resulting larger $x_t$ than that of $z_t>0$ ($r_t>0$). Then according to the GARCH equation, a larger $x_t$ (when $r_t\le0$) will result in a larger $\hat{h}_{t+1}$ as the coefficient $\gamma$ of $\text{log}(x_t)$ is positive. Therefore, such observation is consistent with the well-known leverage effect in stock markets: negative return of day $t$ leads to higher volatility of day $t+1$.


Third, we discuss how the leverage effect can be captured by the proposed threshold measurement equation. As in Table \ref{parameter_estimates}, in the estimated Realized-T-M-GARCH and Realized-D-T-GARCH frameworks both $\xi_1$ and $\xi_2$ estimates are negative and significantly different to 0, as their 95\% CIs do not include 0. Meanwhile, we observe that the intercept term $\xi_1$ estimate in the $r_{t-1}\le0$ regime is larger than the $\xi_2$ estimate in the $r_{t-1}>0$ regime. Such observations are consistent with observations from the Realized-GARCH model as discussed above, e.g., a larger $\xi+\tau_1 z_t$ term when $r_t\le0$ than $r_t>0$. These observations are also in line with the estimates of the intercept terms $\omega_1$ and $\omega_2$ in the GARCH equations of Realized-Threshold-GARCH and the proposed Realized-Double-Threshold-GARCH, i.e.,  the estimated intercept term $\omega_1$ in the $r_{t-1}\le0$ regime is larger than $\omega_2$ in the $r_{t-1}>0$ regime.


We further illustrate how the leverage effect is successfully captured by the proposed threshold measurement equation, using the estimated Realized-T-M-GARCH model as example. We set the value of $\hat{h}_t$ as 1.2074, which is equal to the mean of the in-sample $h_t$ estimates with the first moving window of S\&P 500 returns. The value of measurement equation error term $\varepsilon_t$ is set to 0, as its zero mean assumption. Based on these, for the two regimes $r_t\le0$ and $r_t>0$ the corresponding values of $x_t$ are 0.9227 and 0.7778 respectively. Such observation is in line with the one from the Realized-GARCH measurement equation discussed above:  $x_t$ will be larger when $r_t\le0$ than $r_t>0$. Lastly, the conditional variance forecasts $\hat{h}_{t+1}$ are 1.2307 ($r_{t-1}\le0$ regime) and 1.1689 ($r_{t-1}>0$ regime) respectively. The results demonstrate the leverage effect: negative return of of day $t$ leads to higher volatility of day $t+1$. In addition, one may argue the selected values of $\hat{h}_t$ might affect how/whether the leverage effect can be captured, as in the proposed measurement equation $\hat{\varphi}_1 < \hat{\varphi}_2 $. We have done comprehensive testing on this. In this example, any reasonable choice of $\hat{h}_t$, e.g., even value equal to 5 times of the maximum value of in-sample $h_t$ estimates, still produces $\hat{h}_{t+1}$ value that is larger in the $r_t\le0$ regime.





Lastly, compared to the original measurement equation in the Realized-GARCH which has one regression coefficient $\varphi$ that models the observed realized measure $x_t$ and hidden conditional variance $h_t$, the proposed threshold measurement equation includes two separate coefficients $\varphi_1$ and $\varphi_2$ in two regimes, which can model the relationship between $x_t$ and $h_t$ more flexibly.

\subsubsection{Full out-of-sample results}

To further demonstrate how the proposed threshold measurement equation works, Figure \ref{parameter_estimates} shows all the S\&P500 $\xi_1$\&$\xi_2$ and $\varphi_1$\&$\varphi_2$ parameter estimates for the full out-of-sample period (2000 forecasting steps) in Realized-T-M-GARCH-tG.


As in the second plot in Figure \ref{parameter_estimates}, the $\xi_1$ estimates are consistently larger ($r_{t-1} \le 0$ regime) than the $\xi_2$ estimates ($r_{t-1}> 0$ regime). In the third plot, we see that both $\varphi_1$ and $\varphi_2$ parameters are estimated to be close to 1, while different values are observed at different forecasting steps, which introduces further flexibility and could benefit the volatility and risk forecasting accuracy. Lastly, as in the fourth plot in Figure \ref{parameter_estimates}, the persistence level of $\beta+\gamma \varphi_1$ ($r_{t-1} \le 0$ regime) is consistently smaller than $\beta+\gamma \varphi_2$ ($r_{t-1} > 0$ regime). All the above observations illustrate the distinctive behaviours of the model for the $r_{t-1} \le 0$ and $r_{t-1} > 0$ regimes during the out-of-sample period. In the following section, we present further empirical evidence to support the effectiveness of the proposed threshold measurement equation in risk forecasting.


\begin{figure}[htp]
     \centering
\includegraphics[width=\textwidth]{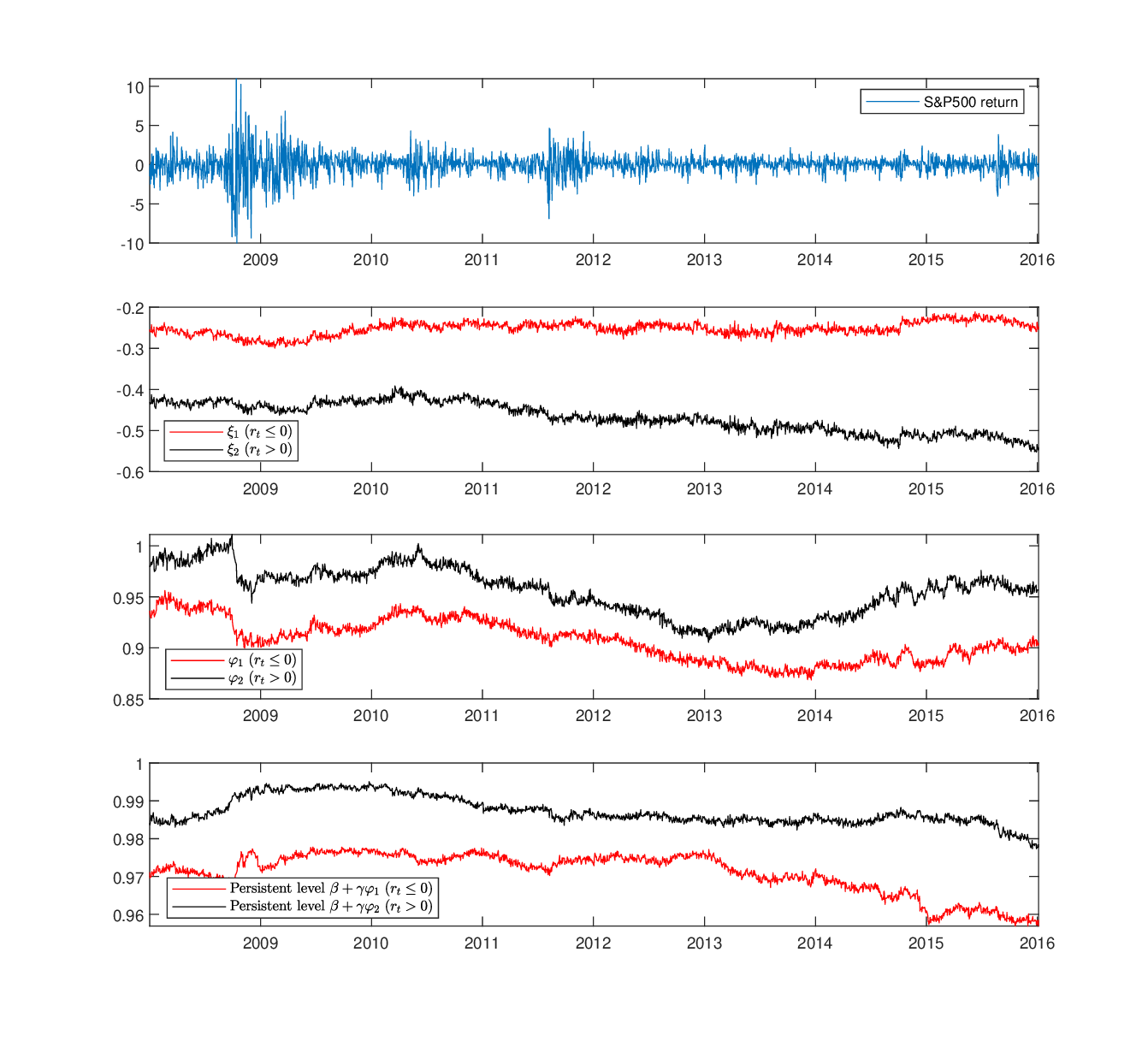}
\caption{\label{parameter_estimates} S\&P 500 out-of-sample threshold measurement equation parameter estimates and persistent levels in Realized-T-M-GARCH-tG.}
\end{figure}



\subsection{ VaR forecasting}
\noindent
Since quantiles are elicitable, as defined in \cite{gneiting2011making}, and the standard quantile loss function is strictly consistent, the expected quantile loss will be a minimum at the true quantile series. In this section, the quantile loss over the out-of-sample period is used to compare the VaR forecast accuracy of the competing models. The most accurate VaR forecasting model is expected to produce the minimized aggregated quantile loss function values, given as in Equation (\ref{q_loss_score}):

\begin{equation}\label{q_loss_score}
\sum_{t=n+1}^{n+m}(\alpha-I(r_t \le \widehat{Q}_t))(r_t-\widehat{Q}_t)  \,\, ,
\end{equation}
where $n$ is the in-sample size and $m$ is the out-of-sample size. $\widehat{Q}_{n+1},\ldots,\widehat{Q}_{n+m}$ is a series of quantile forecasts at levels $\alpha=1\%; 2.5\%$ for the returns $r_{n+1},\ldots,r_{n+m}$.

Values of the out-of-sample quantile loss are reported in Tables \ref{t:quantile_loss_gfc} and \ref{t:quantile_loss}. The average loss across six indices is included in the Avg Loss column. The average rank based on ranks of quantile loss of different models across six markets is calculated and shown in the Avg Rank column. A box indicates the favoured model and a dashed box indicates the 2nd ranked model, based on the average loss and average rank.

For the GFC forecasting study, Table \ref{t:quantile_loss_gfc} shows that the proposed Realized-T-M-GARCH-tG and Realized-D-T-GARCH-tG frameworks are characterized by very competitive performance, on both 1\% and 2.5\% probability levels. On the 1\% level, Realized-D-T-GARCH-tG ranks as the best on average (2.50). The Realized-T-M-GARCH-tG produces the smallest average loss (25.3) and has the second best average rank of 2.83. On the 2.5\% level, the best ranked model (3.17) and the model with smallest average quantile loss (52.4) are again Realized-D-T-GARCH-tG and Realized-T-M-GARCH-tG respectively. Lastly, the effectiveness of incorporating the realized measure in the quantile forecasting is clear, with the Realized-GARCH type models in general better ranked with smaller loss produced than the GARCH type models.

Table \ref{t:quantile_loss} includes the quantile forecasting results for the eight year out-of-sample study (out-of-sample size $m=2000$). For both the 1\% and 2.5\% probability levels, the first and second best performing models are Realized-D-T-GARCH-tG and Realized-T-GARCH-tG.

To conclude, for both forecasting studies on 1\% and 2.5\% probability levels, the proposed Realized-D-T-GARCH-tG and Realized-T-M-GARCH-tG models are characterized by very competitive quantile loss results, in comparison to other models. This demonstrates the validity and effectiveness of incorporating the threshold measurement equation to consider the leverage effect and forecast VaR.

\begin{table}[!ht]
\begin{center}
\caption{\label{t:quantile_loss_gfc} \small For the GFC study, quantile loss function values across six indices. Out-of-sample size $m=400$. $\alpha= 1\%; 2.5\%$ .}\tabcolsep=10pt
\tiny
\begin{tabular}{lcccccc|cc} \hline
Model&S\&P500&NASDAQ&FTSE&DAX&SMI&ASX200&Avg Loss&Avg Rank\\
\hline
$\alpha= 1\%$ \\
\hline

EGARCH-t&27.3&32.8&26.1&26.3&26.9&21.5&26.8&5.83\\
GJR-GARCH-t&26.0&32.0&26.8&27.1&27.8&22.3&27.0&6.33\\
EGARCH-t-HS&26.1&32.4&24.8&26.5&26.1&20.7&26.1&4.67\\
GJR-GARCH-t-HS&26.2&32.2&25.7&27.0&27.0&22.0&26.7&6.17\\
Realized-GARCH-tG&26.7&30.0&24.4&26.8&26.3&20.0&25.7&4.00\\
Realized-T-M-GARCH-tG&25.9&29.2&24.4&26.8&25.3&20.1&\fbox{25.3}&\dbox{2.83}\\
Realized-T-GARCH-tG&29.6&32.1&23.4&26.3&25.5&18.7&26.0&3.67\\
Realized-D-T-GARCH-tG&29.3&31.4&23.5&26.3&24.4&18.6&\dbox{25.6}&\fbox{2.50}\\

\hline
$\alpha= 2.5\%$ \\
\hline

EGARCH-t&59.8&62.8&54.7&53.4&51.4&47.8&55.0&6.00\\
GJR-GARCH-t&55.4&62.0&54.9&54.0&52.3&47.2&54.3&5.67\\
EGARCH-t-HS&57.1&62.7&52.3&52.6&50.8&45.5&53.5&4.00\\
GJR-GARCH-t-HS&55.1&62.0&53.1&53.8&52.2&46.1&53.7&4.33\\
Realized-GARCH-tG&56.2&59.4&51.2&55.3&52.6&42.6&52.9&4.83\\
Realized-T-M-GARCH-tG&55.3&58.8&51.0&55.3&51.7&42.7&\fbox{52.4}&\dbox{3.50}\\
Realized-T-GARCH-tG&58.2&62.1&50.1&54.8&51.9&41.7&53.1&4.50\\
Realized-D-T-GARCH-tG&58.0&61.2&50.1&54.7&50.8&41.9&\dbox{52.8}&\fbox{3.17}\\

\hline
Out-of-sample $m$&400&400&400&400&400&400&&\\
In-sample $n$&1905&	1892&	1943&	1936&	1930&1871&&\\
\hline
\end{tabular}
\end{center}
\emph{Note}:\small  Based on average loss and average rank, the box indicates the favoured model, the dashed box indicates the 2nd ranked model.
\end{table}

\begin{table}[!ht]
\begin{center}
\caption{\label{t:quantile_loss} \small  Quantile loss function values across six indices. Out-of-sample size $m=2000$. $\alpha= 1\%; 2.5\%$.}\tabcolsep=10pt
\tiny
\begin{tabular}{lcccccc|cc} \hline
Model&S\&P500&NASDAQ&FTSE&DAX&SMI&ASX200&Avg Loss&Avg Rank\\
\hline
$\alpha= 1\%$ \\
\hline
EGARCH-t&77.4&88.8&72.5&86.3&79.9&64.3&78.2&6.33\\
GJR-GARCH-t&74.9&86.6&73.3&87.8&82.6&64.8&78.3&6.67\\
EGARCH-t-HS&76.7&88.3&71.3&86.9&78.9&64.0&77.7&5.00\\
GJR-GARCH-t-HS&75.2&86.5&71.9&88.0&81.7&64.6&78.0&5.83\\
Realized-GARCH-tG&74.9&84.5&72.2&84.3&80.8&62.9&76.6&3.83\\
Realized-T-M-GARCH-tG&74.5&83.9&72.1&84.4&80.2&63.0&76.3&3.33\\
Realized-T-GARCH-tG&75.6&85.2&70.2&83.8&79.2&61.2&\dbox{75.9}&\dbox{3.00}\\
Realized-D-T-GARCH-tG&75.5&84.4&70.2&84.2&78.9&61.1&\fbox{75.7}&\fbox{2.00}\\

\hline
$\alpha= 2.5\%$ \\
\hline

EGARCH-t&166.9&184.0&155.2&181.6&159.3&140.3&164.6&6.00\\
GJR-GARCH-t&162.9&183.1&156.4&182.9&159.4&141.7&164.4&6.67\\
EGARCH-t-HS&163.7&182.3&153.0&179.5&157.5&138.4&162.4&3.83\\
GJR-GARCH-t-HS&161.0&180.6&154.5&180.7&159.3&139.8&162.6&4.33\\
Realized-GARCH-tG&162.0&178.7&154.1&185.1&161.8&134.9&162.8&5.17\\
Realized-T-M-GARCH-tG&161.2&178.3&153.9&185.5&161.1&135.0&162.5&4.67\\
Realized-T-GARCH-tG&160.4&178.9&151.4&181.9&160.0&133.0&\dbox{160.9}&\dbox{3.00}\\
Realized-D-T-GARCH-tG&160.5&178.3&151.4&182.2&159.3&133.1&\fbox{160.8}&\fbox{2.33}\\

\hline
Out-of-sample $m$&2000&2000&2000&2000&2000&2000&&\\
In-sample $n$&1905&	1892&	1943&	1936&	1930&1871&&\\

\hline
\end{tabular}
\end{center}
\emph{Note}: \small Based on average loss and average rank, the box indicates the favoured model, the dashed box indicates the 2nd ranked model.
\end{table}

\subsection{ES forecasting}
\noindent
The same set of models is employed to generate one-step-ahead forecasts of 1\% and 2.5\% ES during the forecast period for all six series.

To evaluate the proposed Realized-T-M-GARCH, we assess the performance of different models under comparison to forecast VaR and ES jointly, employing a strictly consistent VaR and ES joint loss function.

\cite{Fissler2016} find the class of jointly consistent scoring functions for VaR and ES, that is, their expectations are uniquely minimized by the true VaR and ES series. The general form of this functional family is:
\begin{eqnarray}
S_t(r_t, Q_t, \text{ES}_t) &=& (I(r_t \le Q_t) -\alpha)G_1(Q_t) - I(r_t \le Q_t) G_1(r_t)  \\ \nonumber
&+& G_2(\text{ES}_t)\left(\text{ES}_t-Q_t + I(r_t \le Q_t)\frac{Q_t-r_t}{\alpha}\right) \\ \nonumber
                      &-& H(\text{ES}_t) + a(r_t) \, ,
                      \label{e:fzloss}
\end{eqnarray}
where $G_1(.)$ is increasing, $G_2(.)$ is strictly increasing and strictly convex,
$G_2 = H^{'}$ and $\lim_{x\to -\infty} G_2(x) = 0$ and $a(\cdot)$ is a real-valued integrable function.

As presented in \cite{tayl2017}, assuming $r_t$ to have zero mean, making the choices: $G_1(x) =0$,
$G_2(x) = -1/x$, $H(x)= -\text{log}(-x)$ and  $a= 1-\text{log} (1-\alpha)$, which satisfy the required criteria, returns the scoring function:
\begin{eqnarray}\label{es_caviar_log_score}
S_t(r_t, Q_t, \text{ES}_t) = -\text{log} \left( \frac{\alpha-1}{\text{ES}_t} \right) - {\frac{(r_t-Q_t)(\alpha-I(r_t \le Q_t))}{\alpha \text{ES}_t}}.
\end{eqnarray}
\cite{tayl2017} refers to Equation (\ref{es_caviar_log_score}) as the Asymmetric Laplace (AL) log score which is used to jointly assess VaR and ES forecasting accuracy in our paper.

\begin{figure}[htp]
     \centering
\includegraphics[width= 0.9\textwidth]{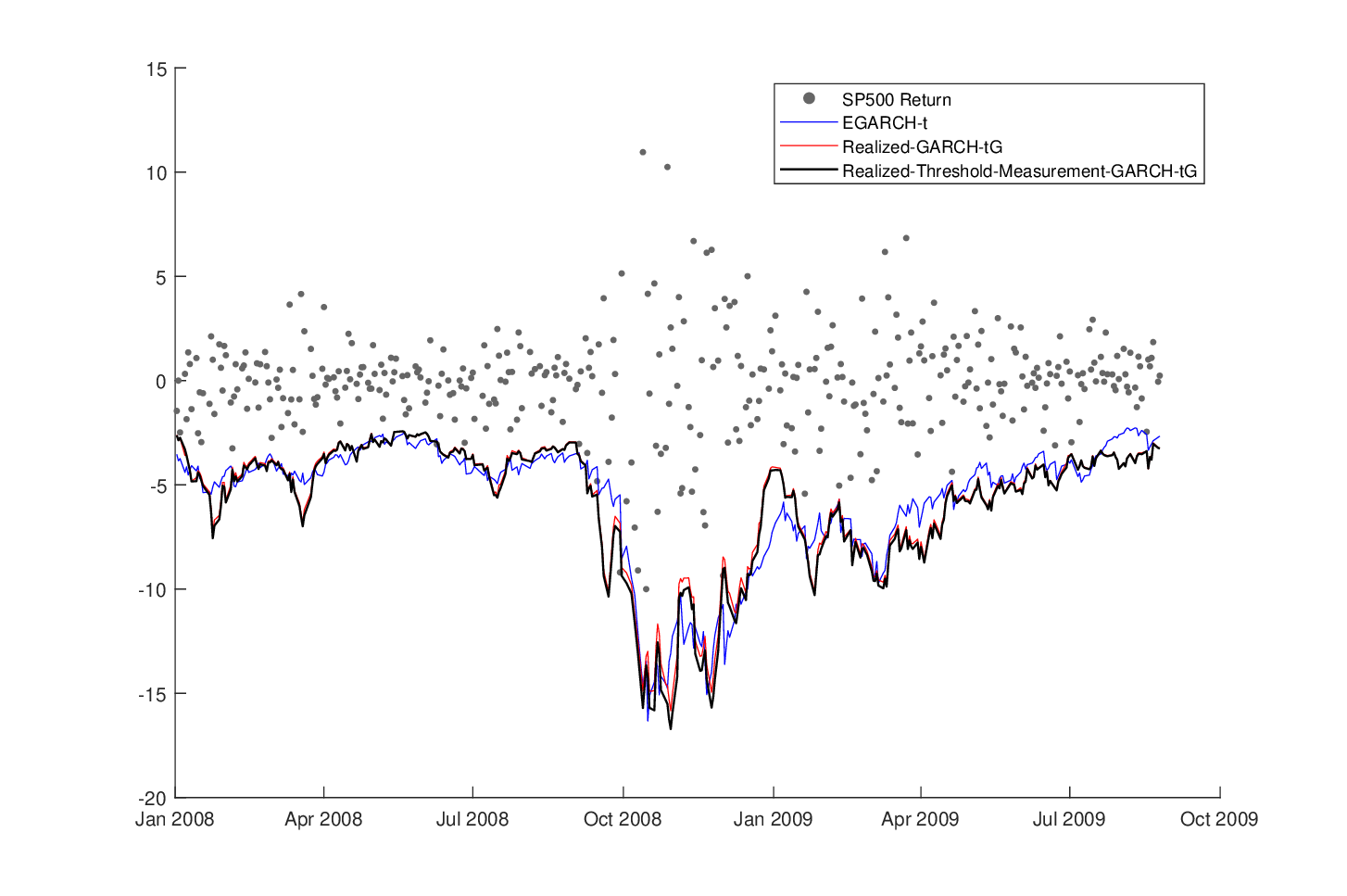}
\caption{\label{es_forecast_plot} S\&P 500 1\% ES forecasts from EGARCH-t, Realized-GARCH-tG, and Realized-T-M-GARCH-tG during the 2008 GFC period.}
\end{figure}

\begin{figure}[htp]
     \centering
\includegraphics[width= 0.9\textwidth]{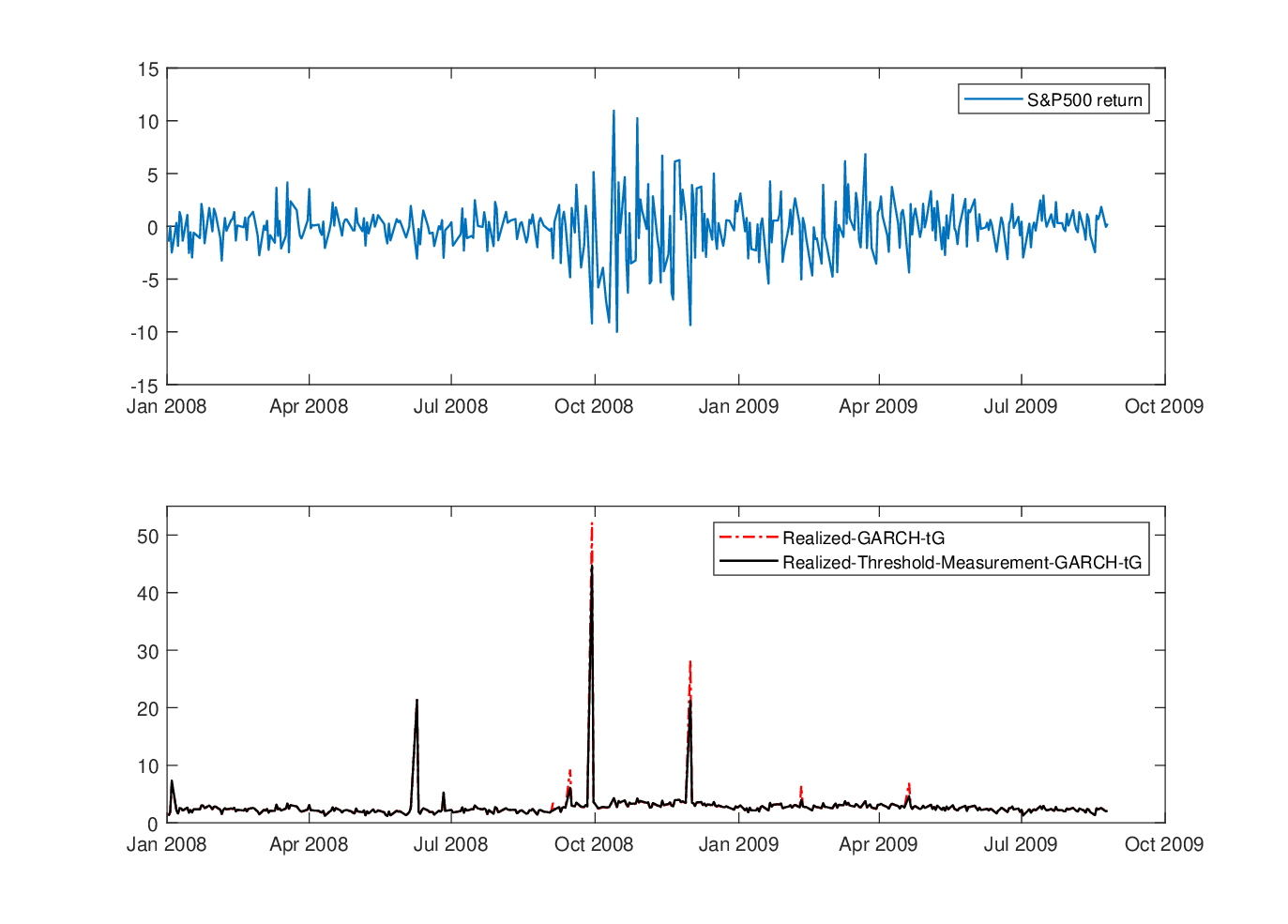}
\caption{\label{jonit_loss_plot} S\&P 500 1\% VaR and ES forecasts AL joint loss values from Realized-GARCH-tG and Realized-T-M-GARCH-tG during the GFC period.}
\end{figure}

First, Figure \ref{es_forecast_plot} shows the S\&P500 1\% ES forecasts from EGARCH-t, Realized-GARCH-tG and Realized-T-M-GARCH-tG during the GFC period. As can be seen, the ES forecasts from Realized-T-M-GARCH-tG and Realized-GARCH-tG present some distinctive behaviours during the highly volatile period, e.g., around October 2008. To make a more in-depth comparison of Realized-GARCH-tG and Realized-T-M-GARCH-tG models, Figure \ref{jonit_loss_plot} presents their S\&P 500 1\% VaR and ES AL joint loss (log-score) values for each time step across the out-of-sample period of the GFC study. In general, the VaR and ES forecasts from Realized-T-M-GARCH-tG are characterized by smaller joint loss values than the ones from the Realized-GARCH-tG, for example in the October 2008 and December 2008 periods. In addition, there are two joint loss value jumps for the Realized-GARCH-tG model during the first half of 2009, while such jumps are smaller for the Realized-T-M-GARCH-tG model. Such reduced VaR and ES joint loss values, especially during the high volatility period, reflect the additional tail risk forecasting efficiency that can be gained from employing the threshold measurement specification in a Realized-GARCH framework. These observations are also consistently presented across different indices.

As a more comprehensive comparison, Tables \ref{t:veloss_gfc} and \ref{t:veloss} report, for each model and index, values of the loss function in Equation (\ref{es_caviar_log_score}) aggregated over the out-of-sample period: $\mathbf{S}= \sum_{t=n+1}^{n+m} S_{t}(r_t, \widehat{Q}_t, \widehat{\text{ES}}_t)$, with $m=400$ and $m=2000$ respectively.

Regarding the joint loss in the GFC study as in Table \ref{t:veloss_gfc}, the proposed Realized-D-T-GARCH-tG and Realized-T-M-GARCH-tG are again the best ranked model and the model with the smallest average loss respectively, on both 1\% and 2.5\% probability levels. Comparing models with and without the high frequency information, the performance of Realized-GARCH type models are clearly preferred.

With respect to the joint loss values produced over the longer forecasting horizon as in Table \ref{t:veloss}, on both the 1\% and 2.\% probability levels the preferred model with the smallest average loss and best rank is Realized-T-GARCH-tG which is closely followed by Realized-D-T-GARCH-tG.

Overall, the joint loss results lend further support to the usefulness of the proposed Realized-T-M-GARCH and Realized-D-T-GARCH in risk forecasting. Both models produce competitive results, when compared to the other competing models, for both forecasting studies and both probability levels. In particular, for the highly volatile GFC period the Realized-T-M-GARCH and Realized-D-T-GARCH produce preferred loss results comparing to the Realized-GARCH and Realized-T-GARCH, with potential reasons for the better performance presented in Figure \ref{jonit_loss_plot}.

\begin{table}[!ht]
\begin{center}
\caption{\label{t:veloss_gfc} \small For the GFC study, VaR and ES joint loss function values across six indices. Out-of-sample size $m=400$. $\alpha= 1\%; 2.5\%$.}\tabcolsep=10pt
\tiny
\begin{tabular}{lcccccc|cc} \hline
Model&S\&P500&NASDAQ&FTSE&DAX&SMI&ASX200&Avg Loss&Avg Rank\\
\hline
$\alpha= 1\%$ \\
\hline

EGARCH-t&1167.6&1244.6&1205.5&1161.1&1135.5&1066.2&1163.4&6.17\\
GJR-GARCH-t&1123.9&1228.1&1208.2&1186.8&1170.6&1088.8&1167.7&7.17\\
EGARCH-t-HS&1134.0&1226.3&1160.9&1156.3&1106&1039.6&1137.2&4.67\\
GJR-GARCH-t-HS&1119.8&1226.4&1164.1&1174.5&1139.1&1068.3&1148.7&5.50\\
Realized-GARCH-tG&1120.0&1182.7&1103.3&1176.2&1104.2&1028.0&1119.1&3.67\\
Realized-T-M-GARCH-tG&1104.6&1175.1&1102.1&1178.9&1082.9&1031.1&\fbox{1112.5}&\dbox{3.00}\\
Realized-T-GARCH-tG&1176.1&1206.9&1089.7&1153.6&1095.7&1000.7&1120.5&3.17\\
Realized-D-T-GARCH-tG&1168.3&1197.3&1090.3&1153.9&1077.7&999.2&\dbox{1114.5}&\fbox{2.67}\\

\hline
$\alpha= 2.5\%$ \\
\hline

EGARCH-t&1114.2&1131.5&1108.8&1068.2&1031.9&1030.3&1080.8&6.50\\
GJR-GARCH-t&1078.6&1124.4&1107.0&1082.6&1047.7&1027.9&1078.1&6.33\\
EGARCH-t-HS&1085.1&1126.6&1073.3&1058.4&1019.4&995.1&1059.7&4.33\\
GJR-GARCH-t-HS&1069.3&1122.5&1073.9&1076.9&1037.7&1007.2&1064.6&5.33\\
Realized-GARCH-tG&1067.6&1091.4&1038.6&1088.3&1035.4&975.6&1049.5&4.00\\
Realized-T-M-GARCH-tG&1059.7&1086.5&1037.2&1090.0&1018.9&977.3&\fbox{1044.9}&\dbox{3.17}\\
Realized-T-GARCH-tG&1093.0&1111.4&1031.2&1075.1&1024.1&963.3&1049.7&3.83\\
Realized-D-T-GARCH-tG&1089.8&1103.3&1031.1&1075.0&1014.2&963.3&\dbox{1046.1}&\fbox{2.50}\\

\hline
Out-of-sample $m$&400&400&400&400&400&400&&\\
In-sample $n$&1905&	1892&	1943&	1936&	1930&1871&&\\
\hline
\end{tabular}
\end{center}
\emph{Note}: \small Based on average loss and average rank, the box indicates the favoured model, the dashed box indicates the 2nd ranked model.
\end{table}

\begin{table}[!ht]
\begin{center}
\caption{\label{t:veloss} \small VaR and ES joint loss function values across six indices. Out-of-sample size $m=2000$. $\alpha= 1\%; 2.5\%$.}\tabcolsep=10pt
\tiny
\begin{tabular}{lcccccc|cc} \hline
Model&S\&P500&NASDAQ&FTSE&DAX&SMI&ASX200&Avg Loss&Avg Rank\\
\hline
$\alpha= 1\%$ \\
\hline
EGARCH-t&4587.6&4837.2&4537.7&4949.5&4689.9&4277.6&4646.6&7.67\\
GJR-GARCH-t&4459.0&4743.6&4486.0&4981.8&4778.7&4245.0&4615.7&6.83\\
EGARCH-t-HS&4539.0&4782.7&4479.9&4924.8&4604.2&4258.8&4598.2&5.50\\
GJR-GARCH-t-HS&4436.8&4717.4&4422.8&4947.4&4684.5&4225.1&4572.3&5.33\\
Realized-GARCH-tG&4415.8&4715.0&4420.3&4822.6&4630.5&4149.5&4525.6&3.33\\
Realized-T-M-GARCH-tG&4420.8&4705.3&4416.4&4827.7&4637.9&4152.6&4526.8&3.67\\
Realized-T-GARCH-tG&4411.4&4711.9&4360.9&4790.3&4610.6&4111.7&\fbox{4499.5}&\fbox{1.67}\\
Realized-D-T-GARCH-tG&4414.5&4693.2&4359.7&4808.9&4635.8&4111.7&\dbox{4504.0}&\dbox{2.00}\\

\hline
$\alpha= 2.5\%$ \\
\hline
EGARCH-t&4290.6&4517.6&4192.1&4585.2&4245.2&3990.4&4303.5&7.33\\
GJR-GARCH-t&4239.2&4490.2&4183.8&4601.4&4253.6&4009.1&4296.2&7.50\\
EGARCH-t-HS&4225.1&4464.5&4146.9&4542.8&4185.5&3953.9&4253.1&4.33\\
GJR-GARCH-t-HS&4174.1&4428.8&4131.3&4556.6&4216.0&3962.5&4244.9&4.17\\
Realized-GARCH-tG&4183.9&4429.4&4144.5&4582.9&4193.0&3888.0&4236.9&4.00\\
Realized-T-M-GARCH-tG&4185.8&4423.8&4142.1&4592.2&4223.7&3889.8&4242.9&4.83\\
Realized-T-GARCH-tG&4144.5&4419.8&4100.6&4538.1&4198.8&3855.1&\fbox{4209.5}&\fbox{1.67}\\
Realized-D-T-GARCH-tG&4145.7&4412.6&4100.3&4548.6&4203.6&3855.3&\dbox{4211.0}&\dbox{2.17}\\

\hline
Out-of-sample $m$&2000&2000&2000&2000&2000&2000&&\\
In-sample $n$&1905&	1892&	1943&	1936&	1930&1871&&\\
\hline
\end{tabular}
\end{center}
\emph{Note}: \small Based on average loss and average rank, the box indicates the favoured model, the dashed box indicates the 2nd ranked model.
\end{table}

\section{Conclusion}\label{conclusion_section}
\noindent
In this paper, an innovative threshold measurement equation is proposed and its validity is evaluated in Realized-GARCH framework. Through incorporating a threshold regression specification, the proposed measurement equation is capable of capturing the leverage effect in a manner that is different from the one in Realized-GARCH. The contemporaneous dependence between the observed realized measure and hidden volatility is also successfully modelled in the proposed framework, in a way that is potentially more flexible in comparison to that in the Realized-GARCH. This threshold measurement equation can be employed in various extensions of the Realized-GARCH, such as Realized-Threshold-GARCH, Realized-EGARCH, Realized-HAR-GARCH, Realized-CARE, and so on. In this paper, the effectiveness of incorporating the proposed measurement equation in the Realized-Threshold-GARCH is studied.

The estimation of the proposed models employs an adaptive Bayesian MCMC method, the validity of which is evaluated via a simulation study. In an empirical study with six market indices and two out-of-sample sizes, the effectiveness of the proposed model is evaluated. The proposed threshold measurement equation produces $\xi_1$ and $\xi_2$ estimates that are capable of adjusting the bias dependent on the sign of the return. In addition, the $\varphi_1$ and $\varphi_2$ parameters in the threshold measurement equation present different behaviours that could potentially model the relationship between the realized measure and volatility in a more flexible way, compared to the original measurement equation in Realized-GARCH. How the proposed threshold measurement equation can capture the leverage effect successfully is also carefully examined.

The 1\% and 2.5\% VaR and ES forecasting results lend further evidence to the usefulness of the proposed framework. Compared to the Realized-GARCH and Realized-T-GARCH, the Realized-T-M-GARCH and Realized-D-T-GARCH models produce competitive quantile loss and joint loss values for an eight year out-of-sample study. Focusing on the high volatility GFC period, the Realized-T-M-GARCH and Realized-D-T-GARCH models are in general favoured.

This work could be extended in a number of ways. First, the effectiveness of incorporating the proposed framework in Realized-EGARCH, Realized-HAR-GARCH and Realized-CARE could be evaluated. Second, multiple realized measures could be considered as input to the proposed threshold measurement equation. The impact of an extended framework of this kind on volatility and tail risk forecasting accuracy could be investigated. Third, the current threshold specification could be potentially extended by a smooth transition framework; see the smooth transition GARCH of \cite{gonzalez1998smooth} and \cite{anderson1999asymmetric}, or the smooth transition dynamic range models of \cite{lin2012forecasting} as examples.

\clearpage

\bibliographystyle{chicago}
\bibliography{bibliography}

\end{document}